\documentclass[12pt,preprint]{aastex}

\newcommand{\ha}{H$\alpha$}
\newcommand{\oi}{[\ion{O}{1}]}
\newcommand{\oiw}{[\ion{O}{1}]~$\lambda$6300}
\newcommand{\hi}{\ion{H}{1}}
\newcommand{\kms}{km~s$^{-1}$}

\shorttitle{\ha\ toward HD~93521 and the Lockman Window}
\shortauthors{Hausen et al.}

\begin{document}
\title{Interstellar \ha\ Line Profiles toward HD~93521 and the Lockman Window}

\author{N. R. Hausen, R. J. Reynolds, L. M. Haffner}
\affil{Department of Astronomy, University of Wisconsin--Madison} 
\affil{475 North Charter Street, Madison, WI 53706}
\email{hausen@astro.wisc.edu, reynolds@astro.wisc.edu, haffner@astro.wisc.edu}
\and
\author{S. L. Tufte}
\affil{Department of Physics, Lewis \& Clark College}
\affil{0615 SW Palatine Hill Road, Portland, OR 97219}
\email{tufte@lclark.edu}

\begin{abstract}
  
  We have used the Wisconsin \ha\ Mapper (WHAM) facility to measure
  the interstellar \ha\ emission toward the high Galactic latitude O
  star HD~93521 ($\ell~=~183\fdg1$, $b~=~+62\fdg2$).  Three emission
  components were detected along this line of sight.  These components
  have radial velocities of --10~\kms, --51~\kms, and --90~\kms\ with
  respect to the local standard of rest (LSR) and have \ha\ 
  intensities of 0.20~R, 0.15~R, and 0.023~R, respectively,
  corresponding to emission measures of 0.55 cm$^{-6}$ pc, 0.42
  cm$^{-6}$ pc, and 0.06 cm$^{-6}$ pc. We have also measured the \ha\ 
  emission toward the direction $\ell~=~148\fdg5$, $b~=~+53\fdg0$,
  which lies in the region of exceptionally low \hi\ column density
  known as the Lockman Window.  An emission component is detected in
  this direction at --1~\kms\ (LSR) with an intensity of 0.20~R (0.55
  cm$^{-6}$ pc). In addition, we studied the direction
  $\ell~=~163\fdg5$, $b~=~+53\fdg5$. No Galactic emission was
  detected along this line of sight, and upper limits on the possible
  intensity of Galactic emission toward this direction are 0.11~R at
  the LSR and 0.06~R at --50~\kms. As part of the process of
  separating the interstellar from the terrestrial emission, we also
  detected and characterized twelve faint ($\sim$0.03-0.15~R),
  unidentified atmospheric lines present in WHAM \ha\ spectra. Lastly,
  we have used WHAM to obtain \oiw\ spectra along the line of sight
  toward HD~93521. We do not conclusively detect interstellar \oi\ 
  emission toward the star, but place an upper limit of 0.060~R on the
  \oi\ intensity of the intermediate velocity (--51~\kms) component.
  If the temperature of the gas is 10,000 K, this limit implies that
  within the \ha\ emitting region, the hydrogen ionization fraction
  $n(\mathrm{H}^+)/n(\mathrm{H}_{\mathrm{total}}) > 0.6$.

\end{abstract}


\section{INTRODUCTION}
\label{sec:intro}

The Wisconsin \ha\ Mapper (WHAM) is a dual-etalon, Fabry-Perot
spectrometer designed to measure optical emission from diffuse
interstellar sources. WHAM is approximately 100 times more sensitive than
the previous instrument of its kind, making it possible to study very low
levels of emission not accessible in the past.  The WHAM \ha\ studies
presented in this paper were made along three lines of sight in which the
interstellar \ha\ emission is extremely faint.  In particular, we have
measured the \ha\ emission line profile toward HD~93521
($\ell~=~183\fdg1$, $b~=~+62\fdg2$), supplementing the work of
\cite{sf93}, who obtained ultraviolet \emph{HST} absorption line
measurements for numerous species toward this star. Secondly, we have
measured the \ha\ emission toward $\ell~=~148\fdg5$, $b~=~+53\fdg0$,
which lies in the region of sky known as the Lockman Window.  This region
has the lowest \hi\ column density of any section of the northern sky, and
thorough 21~cm studies of the area have been made by \cite{jahoda90} and
\cite{lockman86}. The \ha\ observations provide information about the
amount of H~II in this unusual direction.  Lastly, we have observed the
line of sight toward $\ell~=~163\fdg5$, $b~=~+53\fdg5$ (referred to below
as ``Off A'').  This direction was selected for investigation based on
WHAM Northern Sky Survey data, which indicated that only minimal levels of
\ha\ emission could be present.  Results for these three directions are
discussed in \S\ref{sec:ha_93521}, \S\ref{sec:lm}, and \S\ref{sec:offA},
respectively.

Accompanying WHAM's ability to detect faint \emph{Galactic} emission is
its ability to detect faint \emph{terrestrial} emission lines as well.  
In fact, WHAM \ha\ spectra centered near the local standard of rest (LSR)
show a total of twelve weak ($\sim$0.03 -- 0.15~R, 1~R~=~10$^{6}$/4$\pi$
photons cm$^{-2}$ s$^{-1}$ sr$^{-1}$) sky lines of unknown origin in
addition to the well known and much brighter geocoronal \ha\ line.  These
weak lines are present at fixed positions with respect to the geocoronal
line, and their intensities vary slightly during the night.  Of these
twelve lines, four on the blue side of the geocoronal line were first
noticed in WHAM spectra used to study high-velocity clouds and are
discussed by \cite{tufte98} and by \cite{tufte97}.  One of the lines to
the red of the geocoronal line is discussed by \cite{haffner98}.  Since
the atmospheric lines can significantly affect measurements of faint
Galactic \ha\ emission, it is important to understand and remove their
contributions to the data.  In \S\ref{sec:atmlines}, we give a
characterization of the twelve faint atmospheric lines based on a study of
a large sample of WHAM \ha\ spectra.

In addition to \ha\, other, fainter emission lines can provide information
about the physical state of the emitting gas.  For example, measurements
of Galactic \oiw\ emission along a given line of sight, when combined with
the \ha\ intensity for that direction, yield information regarding the
ionization state of the emitting gas, a subject of interest and some
controversy for the HD 93521 sightline (e.g., \citealt{sf93, smh00}).  
In 
particular, the hydrogen ionization ratio
$n(\mathrm{H^+})$/$n(\mathrm{H^\circ})$ in the gas can be obtained via the
observed \oi/\ha\ line intensity ratio. The precise relation, as given by
\cite{reynolds98} and references therein, is
\begin{equation}
\frac{I_\mathrm{[O~I]}}{I_{\mathrm{H}\alpha}} =
2.74 \times 10^4 \left(\frac{T_4^{1.854}}{1 + 0.605T_4^{1.105}}\right) 
\mathrm{exp}\left( 
-\frac{2.284}{T_4} \right)
\frac{n(\mathrm{O})}{n(\mathrm{H})}
\left[ \frac{1 + n(\mathrm{H^{\circ}})/n(\mathrm{H^+})}
{1 + (8/9)n(\mathrm{H^{+}})/n(\mathrm{H^\circ})} \right],
\end{equation}
where $I_\mathrm{[O~I]}/I_{\mathrm{H}\alpha}$ is the intensity ratio
measured in energy units and $T_4$ is the electron temperature in units of
10$^4$~K.  The term \emph{n}(O)/\emph{n}(H) is the gas-phase abundance of
oxygen, which we take to be $3.2~\times~10^{-4}$ \citep{mjc98}.  The [O~I]
observations are discussed in \S\ref{sec:oi_93521} and \S\ref{sec:disc}.

\section{OBSERVATIONS} 
\label{sec:obs} 

The Wisconsin \ha\ Mapper facility consists of a 15~cm aperture,
dual-etalon, multichannel Fabry-Perot spectrometer attached to a 0.6~m
siderostat.  WHAM has a 1$\arcdeg$ beam on the sky and 12~\kms\ radial
velocity resolution over a 200~\kms\ spectral range.  This 200~\kms\
interval can be centered on any wavelength between 4800 and 7300~\AA.  A
more detailed description of this instrument and its operation is provided
by \cite{tufte97} and \cite{haffner99}.  WHAM is a remotely-operable
facility located at Kitt Peak National Observatory, and all observations
for this paper were obtained remotely from Madison, WI.

\subsection{\ha}
\label{sec:ha_obs}

An ``on-off'' technique similar to that described by \cite{tufte98} was
used to determine the interstellar \ha\ emission line profiles toward
HD~93521 and the Lockman Window.  With this method, exposures toward the
direction of interest (the ``on'') are alternated with exposures toward an
``off'' direction in which the Galactic emission is significantly weaker.  
The ``off'' spectra are then subtracted from the ``on'' spectra, thereby
removing, or at least significantly reducing, all features common to the
two directions, including the geocoronal \ha\ line, other much fainter
terrestrial emission lines, and the weak \ha\ Fraunhofer absorption line
in the zodiacal light.

A summary of \ha\ observations is given in Table~\ref{table:ha_obs}. Each
``on'' spectrum and each ``off'' spectrum had an exposure time of 600~s,
except the Lockman Window spectra noted in Table~\ref{table:ha_obs}.
HD~93521 is the ``on'' direction in Data Sets 1a,b,c, 3, and 7.  The
Lockman Window is the ``on'' direction in Data Sets 5 and 6. For the
HD~93521 observations that used the Lockman Window as the ``off,'' the
``on-off'' technique was complicated by the presence of a significant
amount of Galactic emission toward the Lockman direction.  Our correction
for this is discussed in \S\ref{sec:ha_93521}.  When ``Off A'' was used
(Data Sets 5, 6, and 7), no correction for Galactic emission was made,
because the Galactic emission in this direction is below the detection
limit of these observations (see \S\ref{sec:offA}).  In addition,
observations were repeated at different times of the year in order to
utilize the shifting positions of the Galactic emission relative to the
fixed positions of the atmospheric lines present in the spectra.  
Consistency between results was used as a check that the detected emission
was Galactic in origin and not the result of contamination from the
atmospheric lines.

\subsection{Faint Atmospheric Lines near \ha}
\label{atmlines_obs}

Our characterization of the faint atmospheric lines is based primarily on
a study of more than eighty 60~s spectra toward the Lockman Window, taken
on roughly 20 nights spanning a 14-month period in 1997 and 1998.  Of
these spectra, approximately half had the geocoronal line positioned on
the red side of WHAM's 200~\kms\ spectral range, while the other half had
the geocoronal line on the blue side of the spectrum.  These data were
gathered routinely during observing sessions for the WHAM Northern Sky 
Survey.  
In addition, we made use of many of the Lockman Window spectra listed in
Table~\ref{table:ha_obs}, including the spectra (each 600~s exposures) in
Data Sets 2a,b and 4a,b.

Obtaining Lockman Window spectra over many months was important to our
being able to separate the Galactic from the terrestrial emission, since
at any one time, the Galactic emission near the LSR blends with one or
more of the atmospheric lines, which have intensities comparable to that
of the interstellar \ha.  Over the course of the year, the position of the
Galactic emission moves relative to the positions of the terrestrial
lines, making the detection of the Galactic emission possible.  In the
process we were also able to characterize the atmospheric lines in the
spectrum.

We also studied numerous high-latitude ``blocks'' taken from the WHAM
Survey data (\citealt{survey}; also see
\texttt{\url{http://www.astro.wisc.edu/wham/}}).  Each block was the average
of about forty nine 1$\arcdeg$ pointings (each with 30~s of exposure
time) covering a roughly 6$\arcdeg$~$\times$~7$\arcdeg$ region of sky.
These blocks contained only low levels of Galactic emission, making it
possible to independently test the results from the Lockman Window
studies.  This assured us that the atmospheric line parameters that we
had determined were not influenced by the Galactic emission toward a
particular direction.

\subsection{\oiw}
\label{sec:oi_obs}

Two bright atmospheric lines, the \oi\ airglow line and an OH line at
6297.9 \AA, are present in the \oiw\ spectra, making the detection of
faint Galactic \oi\ emission quite difficult. The \oiw\ studies for
HD~93521 were therefore also based on the ``on-off'' method. Furthermore,
for comparison purposes, observations were repeated at different times of
year as the position of the Galactic emission shifted with respect to the
terrestrial lines. Table~\ref{table:oi_obs} lists the observations.  The
Lockman Window is the ``off'' direction for Data Sets 8a,b, 9a,b, and
10a,b, whereas Data Set 11 uses Off A as the ``off.'' With the exception
of the Lockman Window spectra noted in Table~\ref{table:oi_obs}, all
individual spectra were 600~s exposures. Spectra in Data Set 11 were taken
in an ``on''/``off'' sequence like that described for the \ha\ Data Sets
(see \S\ref{sec:ha_obs}).  Data in the remaining \oi\ Data Sets were
gathered in the repeated sequence ``on,'' ``off,'' ``on,'' yielding twice
as much integration time for the ``on'' direction as for the ``off.''

\section{RESULTS}
\label{sec:results}

\subsection{Faint Atmospheric Lines near \ha}
\label{sec:atmlines}

\begin{figure}[htbp]
  \begin{center}
    \plotone{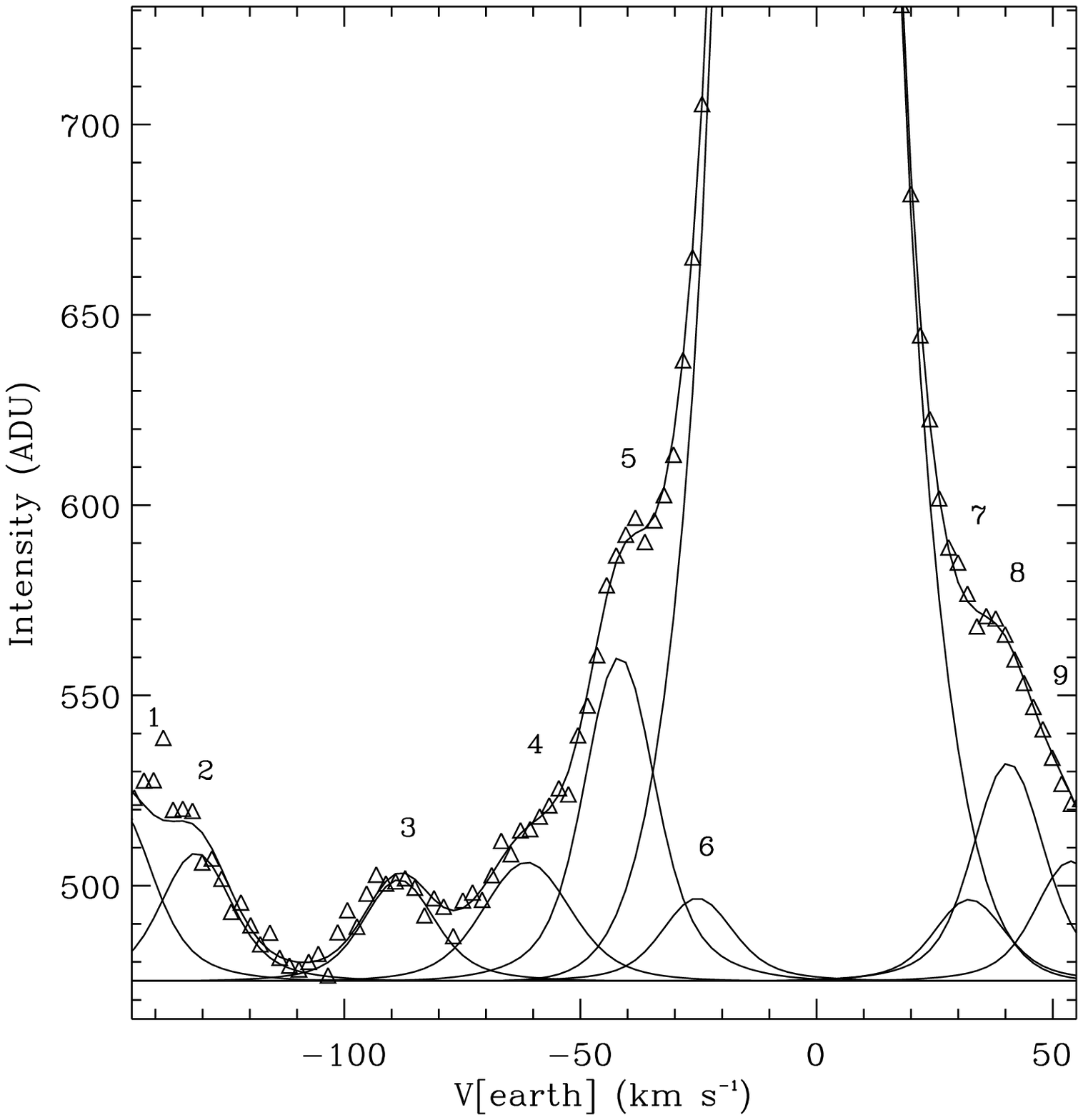}
    \caption{A spectrum of the faint atmospheric lines near \ha,
      with the geocoronal line positioned on the red side of the
      spectrum.  The interstellar emission has been subtracted from
      this spectrum.  The weak atmospheric lines are fit with solid
      lines and are numbered according to the line identifications in
      Table~\ref{table:atmlines}.  The bright geocoronal line at
      --2~\kms\ with respect to the earth is also fit (the shift from
      zero is a fine structure effect, resulting from the fact that
      this line is not a recombination line).  The solid line through
      the data points shows the composite fit.  A 15 \kms\ wide line
      with an intensity of 0.1~R would have a peak height above
      background of about 92 arbitrary data units (ADU) on the
      intensity scale.  The vertical scale has been expanded to show
      the weak atmospheric lines. The peak of the geocoronal line is
      at 4250~ADU.}
    \label{fig:mlines}
  \end{center}
\end{figure}

\begin{figure}[htbp]
  \begin{center}
    \plotone{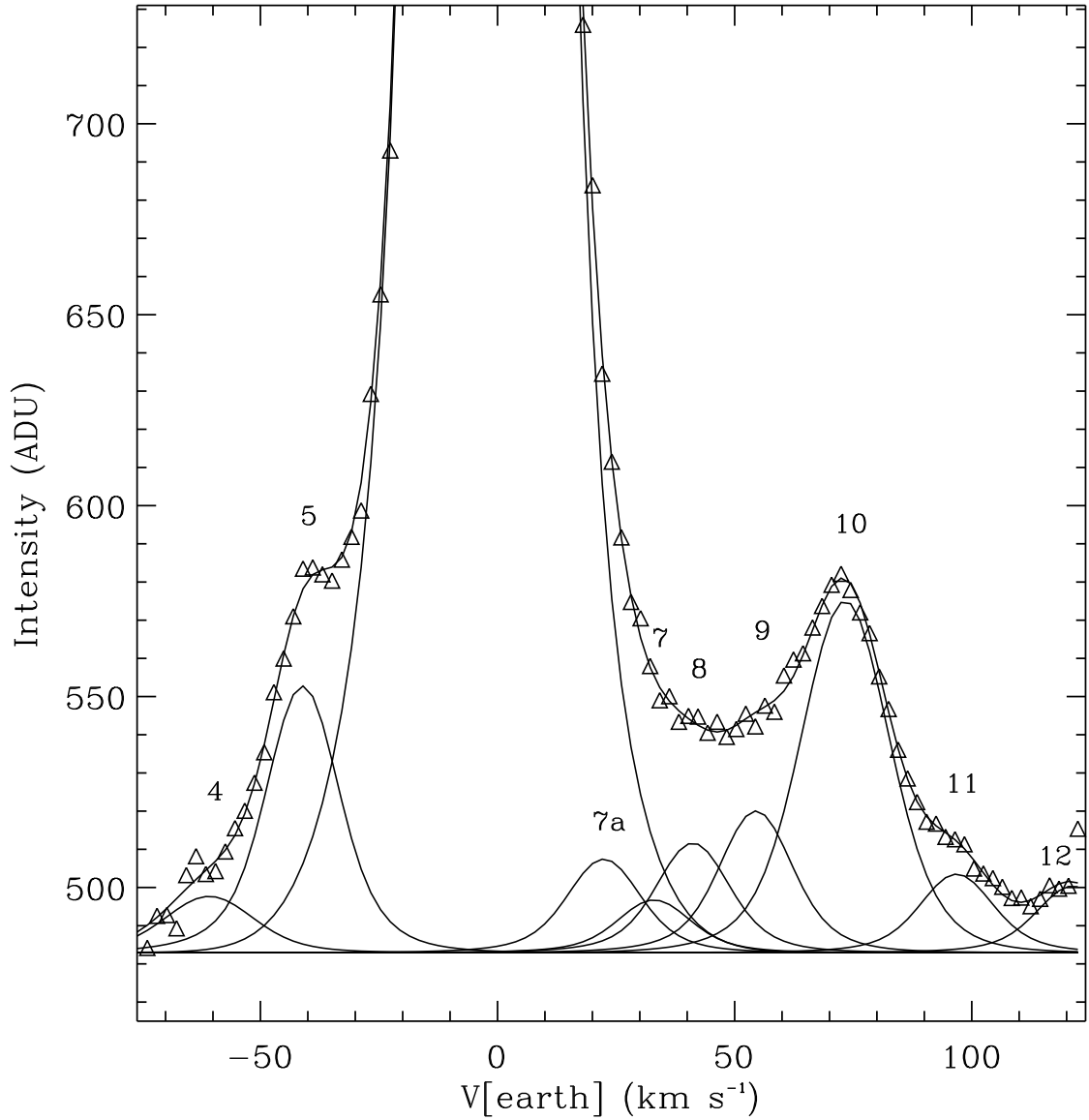}
    \caption{A spectrum of faint atmospheric lines near \ha\
      similar to that in Figure~\ref{fig:mlines}, except the
      geocoronal line is positioned on the blue side of spectrum and
      the observations were taken at a different time of the year. The
      vertical scale matches that of Figure~\ref{fig:mlines}. The peak
      of the geocoronal line is at 3800~ADU.}
    \label{fig:elines}
  \end{center}
\end{figure}

We detected twelve weak atmospheric lines in the WHAM \ha\ spectra
centered near the local standard of rest.  Figure~\ref{fig:mlines},
plotted with respect to the earth's velocity frame, shows the
atmospheric lines that are visible when the geocoronal line is on the
red side of the spectrum.  The spectrum in Figure~\ref{fig:mlines} is
the sum of all Lockman Window spectra in Data Sets 1a,b,c, 3, 4a,b,
and 5, normalized to 600~s. In each of the Data Sets 1, 3, 4, and 5,
the Lockman Window Galactic emission near the LSR (discussed in \S3.3
below) appears at a different geocentric velocity.  Hence, this
Galactic emission was removed from each of the constituent spectra
before summing.  Figure~\ref{fig:elines}, also plotted with respect to
the earth's velocity frame, shows a different wavelength region, with
the geocoronal line located on the blue side of the spectrum.  This
spectrum is the sum of all Lockman Window spectra in Data Sets 2a,b
and 6, normalized to 600~s.  Again, the Galactic emission has been
removed from this spectrum.

Parameters for the faint atmospheric lines are given in
Table~\ref{table:atmlines}.  The low intensity and close spectral
proximity of the lines made it difficult to fit them using free position,
width, and intensity parameters.  A fixed width of 10~\kms\ was therefore
adopted for all lines except those numbered 4 and 10.  The width was fixed
at 15~\kms\ for Lines 4 and 10, which appear broader than the other lines.
Line 10 may in fact be a blend of two components, as some spectra show a
narrower-than-usual line with the position shifted by approximately
0.05~\AA\ from the typical position of Line 10.  It is possible that the
remaining lines are also blends that WHAM does not resolve.  Since the
feature labeled in Figure~\ref{fig:elines} as Line 7a appeared to be
present in only a very limited number of spectra, we did not characterize
it in the same manner as the other atmospheric lines (see
Table~\ref{table:atmlines}).  We have not identified the atomic or
molecular source of any of the weak atmospheric lines.

Table~\ref{table:atmlines} gives the intensity of the lines as shown in
Figures~\ref{fig:mlines} and \ref{fig:elines}.  The lines that are common
to both Figure~\ref{fig:mlines} and Figure~\ref{fig:elines} illustrate the
degree to which the relative intensities of the lines can vary.  For
example, Line 6 is not detected in Figure~\ref{fig:elines}, and Line 7a is
not detected in Figure~\ref{fig:mlines}.  We have found that, in general,
absolute intensities are quite stable ($< 0.05$ R) over periods of a half
hour or more.  Furthermore, the relative intensities between lines are
consistent enough for us to use a set of ``adopted relative intensities'',
and these are given in Table~\ref{table:atmlines}.  In a given spectrum,
the adopted intensity for each line is a percentage of the intensity of
Line 5 in that spectrum. The intensity of Line 5 itself is obtained from a
Gaussian fit to the spectrum.

\subsection{The ``Off A'' Direction}
\label{sec:offA}

During reduction of the WHAM Northern Sky Survey, the Off A direction
($\ell~=~163\fdg5$, $b~=~+53\fdg5$) was identified as having an
exceptionally low level of \ha\ emission.  Subsequent long exposures
toward this direction have confirmed this (Data Set 5, in which the LSR is
maximally to the blue of the geocoronal line for this direction). Although
we detected no Galactic \ha\ emission toward this direction, we are
ultimately limited by the presence of the geocoronal line and by
uncertainties in the intensities of the faint atmospheric lines.  
Table~\ref{table:ha_results} lists the resulting upper limits placed on
possible Galactic emission toward the Off~A direction from 0 to --50~\kms.  
We derived these limits using Data Set 5 spectra and assuming that three
sources of uncertainty in our knowledge of the weak atmospheric lines (the
absolute intensity of Line 5, the exact intensity ratios in a given
spectrum between Line 5 and the other lines, and the uncertainty as to the
presence of Line 6) all add positively in the same direction.

\subsection{Interstellar \ha\ Toward the Lockman Window}
\label{sec:lm}

The spectra in Data Set 5 were used to determine the interstellar \ha\ 
emission line
profile for the Lockman Window direction.  Specifically, the six 600~s Off
A spectra in Data Set 5 were averaged and then subtracted from the average
of the six 600~s Lockman Window spectra in the same Data Set.  The
resulting spectrum is shown in Figure~\ref{fig:main}. In this figure, the
subtraction residual of the geocoronal line (at +21~\kms) has been
removed.  Results from fits of the LSR component in this spectrum are
listed in Table~\ref{table:ha_results}.  The error bars in the table
result from baseline uncertainties and the blend of the Galactic emission
with the geocoronal line residual.  For the intensity of the --1~\kms\
component, there is an additional systematic uncertainty (not included in
Table~\ref{table:ha_results}) of +0.11~R, resulting from the upper limit
on Galactic emission toward the Off A direction.  We also derived the
Galactic \ha\ spectrum toward the Lockman Window by subtracting Lockman
Window spectra obtained in November (when the LSR is at --26~\kms\ wrt the
earth) from Lockman Window spectra obtained in April (when the LSR is at
+15~\kms\ wrt the earth).  This difference spectrum greatly reduced the
atmospheric features and revealed
the Galactic emission as negative and positive spectral features near
--26~\kms\ and +15~\kms\, respectively, with respect to the earth.  The
parameters for the Galactic emission derived from this technique were
consistent with those derived using the Off A direction, indicating
that the interstellar emission toward Off A is indeed less than 0.1~R.  

For comparison, Figure~\ref{fig:main} also shows the \hi\ profile toward
the Lockman Window.  This spectrum is from the Leiden/Dwingeloo survey
\citep{hb97}, and is the average of their $0\fdg5$ pointings whose centers
fall within WHAM's 1$\arcdeg$ beam directed at $\ell = 148\fdg5$, $b =
+53\fdg0$. Since the \hi\ spectrum has an obvious component at --54~\kms,
an upper limit in the \ha\ at this velocity was determined.  The limit,
given in Table~\ref{table:ha_results}, is based on fits to Lockman Window
spectra at various times of the year.  For these fits, a component was
fixed in the spectra at --54~\kms\ with a width of 25~\kms.  Different
possible values of atmospheric line intensities were then considered in
order to establish the Galactic emission upper limit listed in
Table~\ref{table:ha_results}.

Lastly, two noticeable features in the Lockman Window \ha\ spectrum should
be mentioned:  the elevated data points near +60~\kms\ and the rise in the
data at the negative edge of the spectrum.  These features are within the
baseline uncertainty of the spectrum and have no apparent correspondence
in \hi.  It is nevertheless possible that one or both represent very
low-intensity Galactic \ha\ emission. If the true baseline is near zero in
the Figure~\ref{fig:main} spectrum, then the positive-velocity feature is
centered at about +56~\kms\ with an intensity of 0.04~R. The spectra in
Data Set 6 were obtained in an attempt to verify whether this is a real
Galactic emission component.  Unfortunately, because the atmospheric lines
near the region of interest showed marked intensity fluctuations between
temporally-adjacent 600~s spectra in Data Set 6, we were unable to verify
the reality of this weak feature.

No additional data were obtained to test whether the feature at the
negative edge of the spectrum is indeed a high-velocity Galactic emission
component.  However, while using the Lockman Window as an ``off''
direction during a study of emission from high-velocity clouds,
\cite{tufte97} found evidence for the presence of a very weak emission
component toward the Lockman Window near --130~\kms.  This lends support
to the idea that the feature in Figure~\ref{fig:main} is Galactic in
origin, but clearly further ``on'' minus ``off'' observations centered at
a high, negative velocity would be needed to provide a definitive result.

\subsection{Interstellar \ha\ Toward HD 93521}
\label{sec:ha_93521}

``On-off'' studies of the spectra in Data Sets 1a,b,c and 3 reveal three
interstellar \ha\ emission components toward HD~93521.  There is a
``slow'' component at --10~\kms, an ``intermediate velocity'' component at
--51~\kms, and a fainter, ``high velocity'' component at --90~\kms.  Full
results are given in Table~\ref{table:ha_results}.  The error bars given
in Table~\ref{table:ha_results} for HD~93521 incorporate uncertainties
associated with the baseline level and the blend of the slow component
with the residual of the incompletely-subtracted geocoronal line.  They
also include the uncertainty generated by the error bars of the --1~\kms\
Lockman emission.  The results in Table~\ref{table:ha_results} do not
include a systematic uncertainty of +0.11~R at the LSR, due to the upper
limit on the Galactic emission toward Off A, or a systematic uncertainty
of +0.06~R at --50~\kms, due to the upper limit on the Galactic emission
at --50~\kms\ toward the Lockman Window.

The analysis conducted to obtain the results presented in
Table~\ref{table:ha_results} proceeded as follows:  The five 600~s Lockman
Window spectra in Data Sets 1a,b,c were averaged together, and the
Galactic emission at --1~\kms\ (parameters given in
Table~\ref{table:ha_results}) was removed from the average spectrum.  It
was assumed that there was no other Galactic emission in the Lockman
Window spectra.  The corrected average ``off'' spectrum was then
subtracted from the average of the five 600~s HD~93521 spectra in Data
Sets 1a,b,c.  A similar procedure was followed for the data in Data Set 3,
with the additional step of normalizing the average Lockman Window
spectrum to 600~s of exposure time.  Both ``on-off'' spectra were then
studied, although parameters for the slow emission component were
determined from the data in Data Sets 1a,b,c only, as the slow
component was farthest from the incompletely-subtracted geocoronal line in
this case.  During the Data Set 3 analysis, the parameters of the slow
component were fixed at the best-fit values derived from Data Sets 1a,b,c.

As a check on these results, the spectra in Data Set 7 (for which Off A is
the ``off'' direction) were also analyzed using the ``on-off'' method.  
This work was carried out under the assumption that there was no Galactic
emission present in the Off A spectra.  Also, because the Galactic
emission near the LSR was blended with the incompletely-subtracted
geocoronal line, parameters for the slow emission component were fixed
during the analysis at the best-fit values given for this component in
Table~\ref{table:ha_results}.  The intermediate velocity emission
component was then detected at --53~\kms, with a width of 31~\kms\ and an
intensity of 0.13~R.  The high velocity emission component was detected at
--86~\kms, with a width of 20~\kms\ and an intensity of 0.022~R.  These
results are generally consistent with the uncertainties listed in
Table~\ref{table:ha_results}, except for the best-fit width of the
intermediate velocity component, which is slightly outside the error bars
of the value derived from the study that used the Lockman Window as the
``off.'' It is possible that this discrepancy is due to very low levels of
undetected Galactic emission in the Lockman Window and/or Off A
directions. Such emission may produce effects of this magnitude in the
``on-off'' line profile of an ``on'' direction in which the Galactic
emission is faint.

Figure~\ref{fig:main} shows the \ha\ profile toward HD~93521.  This
profile was obtained from the spectra in Data Sets 1a,b,c and 3.  
Specifically, the residuals from the incompletely-subtracted geocoronal
line were removed from the two ``on-off'' spectra generated from these
Data Sets.  (These residuals were located at +26~\kms\ in the spectrum
from Data Sets 1a,b,c and at +14~\kms\ in the Data Set 3 spectrum.) Then
the two resulting spectra were averaged, yielding the spectrum given in
Figure~\ref{fig:main}. Due to the LSR velocity off-set between 1997
November (Data Sets 1a,b,c)  and 1999 January (Data Set 3), several data
points at the positive velocity extreme in the November data and at the
negative velocity extreme in the January data were unmatched with points
in the other spectrum. These points were excluded from the average
spectrum, and hence the velocity scale of the HD~93521 \ha\ spectrum in
Figure~\ref{fig:main} covers only a 188~\kms\ (rather than 200~\kms)
interval.

Figure~\ref{fig:main} shows the \hi\ spectrum toward HD~93521 as well. The
\hi\ spectrum is the average of the Leiden/Dwingeloo pointings whose
centers fall within WHAM's 1$\arcdeg$ beam directed at $\ell = 183\fdg1$,
$b = +62\fdg2$.  The \hi\ spectrum has the same 188 \kms\ velocity scale
as the HD~93521 \ha\ spectrum.

\subsection{A Search for Interstellar \oiw\ Emission Toward HD 93521}
\label{sec:oi_93521}

The strength of the \oi\ emission relative to \ha\ is a measure of the
hydrogen ionization fraction in the emitting gas and thus has the
potential to test the conclusion of \cite{sf93} that the warm ionized gas
toward HD 93521 is mixed with the H~I, forming partially ionized (30\%)
clouds.  Unfortunately, because it is so much weaker than the \ha\, we
were unable to make a clear detection of Galactic \oi\ emission toward
HD~93521, despite using an ``on-off'' technique and repeating observations
at different times of year.  The incomplete subtraction of the very bright
\oi\ airglow line in each ``on-off'' spectrum prevented the detection of
interstellar \oi\ emission near the LSR.  Although the intermediate
velocity component near --50~\kms\ was adequately separated from both the
airglow line and the prominent OH feature 115~\kms\ to the blue of the
airglow line, the residual atmospheric line contamination in the
``on-off'' spectra led to baseline uncertainties that were comparable to
or greater than the intensity of any Galactic emission present near
--50~\kms.

\begin{figure}[htbp]
  \begin{center}
    \plotone{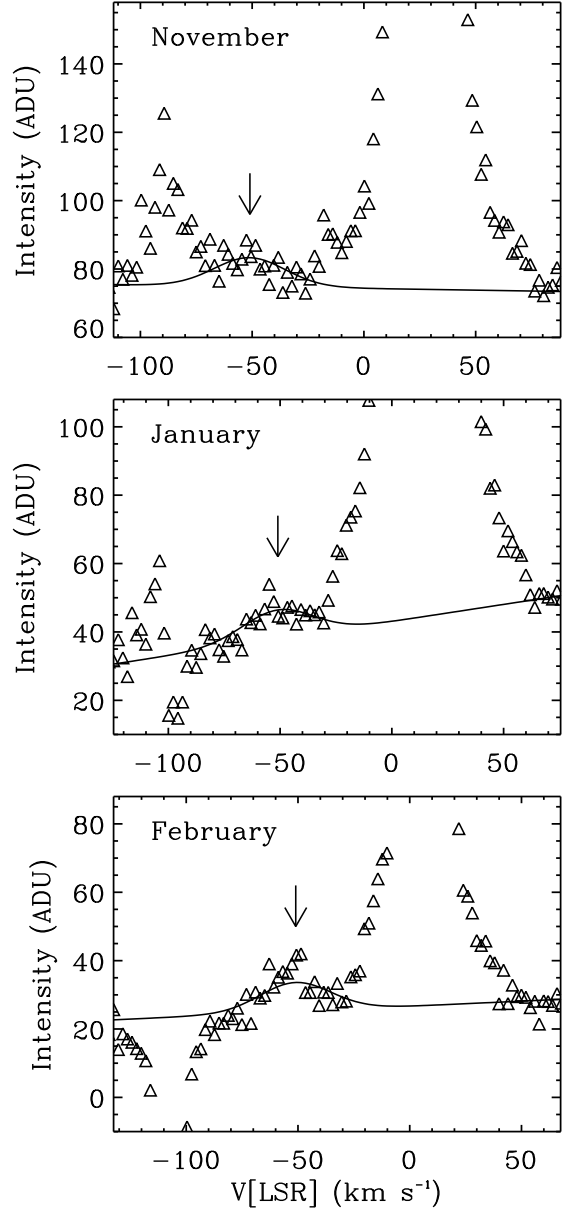}
    \caption{\oiw\ spectra toward HD~93521.  The spectra are
      ``on-off'' subtractions made from data gathered at three
      different times of year (details given in text).  The Gaussian
      fit (solid line) through the data points in each panel is
      centered at the radial velocity of the corresponding \ha\ 
      emission (marked with an arrow). This Gaussian has a width of
      34~\kms\, and an intensity of 0.026~R (see text).  A 34 \kms\ 
      wide line with an intensity of 0.1~R would have a peak height
      above the baseline of about 30 ADUs on this intensity scale.  In
      all three panels, the vertical scale has been expanded.  In the
      top panel, the incompletely-subtracted \oi\ airglow line, at
      +28~\kms (LSR), has a peak height of 1450~ADU.  In the middle
      panel, the airglow line residual is at +14~\kms\ and peaks at
      3240~ADU.  In the bottom panel, the airglow line residual is at
      +~6~\kms\ and peaks at 825~ADU.  The feature near --100~\kms\ in
      the three spectra is the incompletely-subtracted atmospheric OH
      line.}
    \label{fig:oi}
  \end{center}
\end{figure}

Figure~\ref{fig:oi} shows three ``on-off'' \oi\ spectra.  The top panel is
the 600~s average of the ``off'' (the Lockman Window) spectra in Data Sets
8a and 8b subtracted from the 600~s average of the HD~93521 spectra in the
same Data Sets.  The middle and bottom panels show spectra obtained
similarly for Data Sets 9a,b and 11, respectively.  An upper limit on the
possible interstellar \oi\ intensity is given in
Table~\ref{table:oi_results}.  This limit is based on fits to ``on-off''
spectra using data from Data Sets 8a through 11.  For these fits, it was
assumed that the Galactic \oi\ emission was at --51~\kms\ (the velocity of
the \ha\ emission) and had a width of 34~\kms.  This width was
calculated from the width of the \ha\ component at --51~\kms\, assuming
that the \ha\ and \oi-emitting gas is well-mixed and at a temperature of
8000~K.  The Gaussian component shown in each of the spectra in
Figure~\ref{fig:oi} illustrates one possible fit to the data.  This
component has a radial velocity fixed at --51~\kms (LSR), a width fixed at
34~\kms\ (FWHM), and an intensity slightly less than half the upper limit.  
For completeness, Table~\ref{table:oi_results} includes \oi/\ha\ intensity
ratios for both the upper limit and sample fit cases.  These intensity
ratios incorporate a correction for the slight differences in instrument
response (optical transmittance and detector quantum efficiency) between
6300 \AA\ and 6563 \AA.

\section{DISCUSSION AND CONCLUSIONS}
\label{sec:disc}

Interstellar \ha\ emission has been detected and characterized toward two
high Galactic latitude, low H~I column density sightlines, the Lockman
Window ($\ell~=~148\fdg5$, $b~=~+53\fdg0$) and HD 93521
($\ell~=~183\fdg1$, $b~=~+62\fdg2$).  The relatively bright geocoronal
line and the presence of many lower intensity, unidentified atmospheric
emission lines make it difficult to measure accurately the interstellar
\ha\ emission in these faintest parts of the sky.  However, from careful,
long integration WHAM spectra spread over many
months, the interstellar and terrestrial emissions were successfully
separated.  The results provide new constraints on the nature of the
interstellar clouds along these two sightlines and their environment.

\begin{figure}[htbp]
  \begin{center}
    \plotone{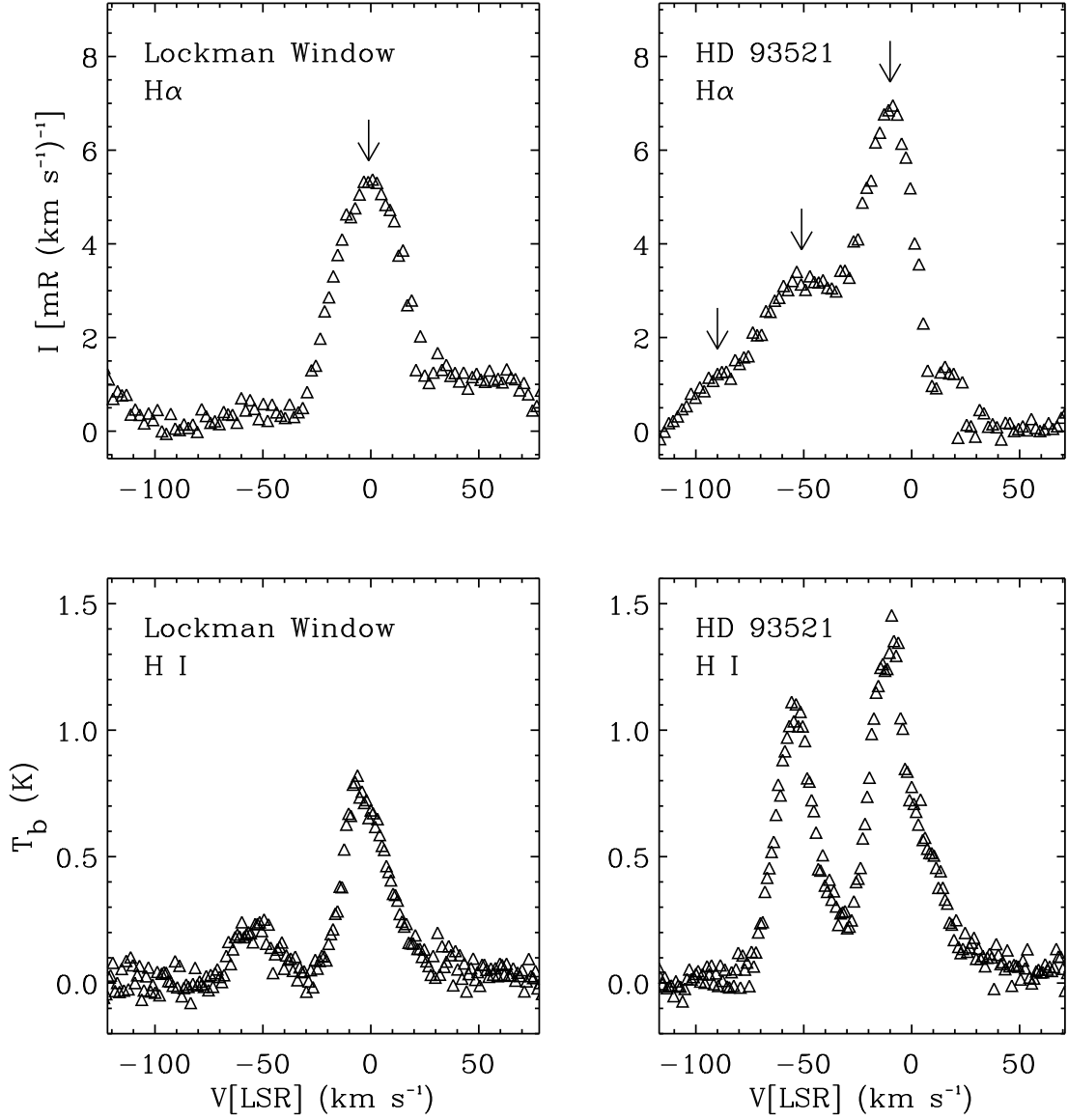}
    \caption{\ha\ and \hi\ spectra toward the Lockman Window and
      HD~93521.  The \ha\ data are ``on-off'' subtraction spectra with
      the residuals from the incompletely-subtracted geocoronal line
      removed (see text).  The intensity scale for the \ha\ spectra is
      in milli-rayleighs (mR) per km~s$^{-1}$.  A constant was
      subtracted from the HD~93521 \ha\ spectrum in order to bring the
      background level to zero.  Arrows mark the velocity of the
      detected emission components.  The \hi\ spectra are from the
      Leiden/Dwingeloo survey \citep{hb97}.}
    \label{fig:main}
  \end{center}
\end{figure}

We have found that the Lockman Window, in addition to having the
lowest H~I column density in the sky ($5 \times 10^{19}$ cm$^{-2}$),
also has unusually weak interstellar \ha\ emission (0.20~R), well
below the 0.8~R average for this Galactic latitude (from the WHAM sky
survey; \citealt{survey}).  This region of the sky thus appears to be
a true low column density window through the Galactic disk, depleted
both in H~I and in H~II.  Only one \ha\ velocity component is detected
toward the Lockman Window, the component near the LSR, even though in
the H~I spectrum there is also a prominent emission component near
--50~\kms\ (see Fig.~\ref{fig:main}).  An \ha\ intensity of 0.20~R for
the Lockman Window implies an emission measure of 0.55 cm$^{-6}$ pc
(at 10$^4$ K), which corresponds to a column density
N$_{\mathrm{H~II}} \approx 2 \times 10^{19}$ cm$^{-2}$, if the mean
density within the ionized regions is about 0.08 cm$^{-3}$ (see
\citealt{reynolds91}).  This is approximately 40\% the H~I column
density and implies that the total hydrogen column density in this
direction is about $7 \times 10^{19}$ cm$^{-2}$.

Toward HD 93521, $21\arcdeg$ away, the slow and intermediate velocity
components detected in \ha\ clearly correspond to the two principal
emission components present in the 21 cm spectrum.  The intensity of the
intermediate velocity (--50~\kms) component relative to the slow
(--10~\kms) component is significantly weaker in the \ha\ spectrum than in
the 21~cm spectrum, suggesting that perhaps an intermediate velocity \ha\
component toward the Lockman Window is just below the detection limit of
the observation.  For HD 93521 there is also strong evidence for a high
velocity \ha\ emission component at --90~\kms, but no corresponding
feature in the H~I spectrum.  This sightline is very near the High
Velocity H~I Cloud Complex M, parts of which have been detected in \ha\ at
radial velocities ranging from --109~\kms\ to --61~\kms\ 
\citep{tufte98}.  
Therefore, the high velocity \ha\ component may be associated with a fully
ionized portion of this high velocity cloud complex.  The total \ha\
intensity (0.37~R) for the HD 93521 sightline corresponds to an emission
measure of 1.0 cm$^{-6}$ pc, and thus N$_{\mathrm{H~II}} \approx 4 \times
10^{19}$ cm$^{-2}$ (about 30\% of the H~I column density of $1.25 \times
10^{20}$ cm$^{-2}$).

Although there is good correspondence between the H~II and H~I for the low
velocity and (for HD 93521) the intermediate velocity components, the
difference in the general appearance of the \ha\ and 21 cm line profiles
strongly suggests that the H~II and the H~I are not mixed together in the
form of partially ionized clouds.  In particular, the \ha\ components
appear systematically broader.  For example, the width (FWHM) of the
intermediate velocity H~I component toward HD 93521 is 20 \kms, only about
half that of the corresponding \ha\ emission.  As a result, there is
significant blending between the slow and intermediate velocity components
in the \ha\ spectrum, whereas these components are well resolved in the
H~I spectrum.  This is not a result of the different spectral resolutions
of the \ha\ and 21~cm observations.  Therefore, rather than being
primarily neutral clouds that are 30\%--40\% ionized, as proposed by
\cite{sf93} and \cite{sciama98}, each component seems to consist of
separate regions of H~I and H~II at nearly the same radial velocity.  
This could be the result of either a close physical association
between the H~I and H~II \citep{mo77} or large scale organized
motions of the interstellar medium along this sightline \citep{wc01}.

Because the \oiw/\ha\ intensity ratio is a probe of the hydrogen
ionization fraction within the emitting gas (see \S\ref{sec:intro}),
observations of \oi\ provide a potentially rigorous test of these
``mixed'' versus ``separated'' models for the H~I and H~II in these
clouds.  Unfortunately, the \oi\ spectrum, which only allows an upper
limit on the intermediate velocity (--50~\kms) component
(Table~\ref{table:oi_results}; see also \S\ref{sec:oi_93521}), does not
provide an unambiguous result.  If the temperature is as low as 6000 K, as
adopted by \cite{sf93} based on the widths of lines generally associated
with H~I clouds, then the \oi\ observations are consistent with their
model of primarily neutral clouds containing a 30\% mixture of H$^+$.  On
the other hand, if the temperature is closer to 10,000 K, as recent
observations of [S~II] $\lambda$6717 and [N~II] $\lambda$6584 suggest
(Pifer et al, in preparation), then these \oi\ results imply that the
emitting regions are predominantly ionized (i.e.,
$n(\mathrm{H}^+)/n(\mathrm{H}_{\mathrm{total}}) > 0.6$).  The \oi\ limit
is also just consistent with recent predictions of a supernova remnant
ionization model for the HD 93521 sightline by \cite{smh00}.  Given the
severe atmospheric line contamination of this region (worse than at \ha ),
it will be extremely difficult with ground-based observations to push the
sensitivity limit of the \oi\ observations down to the level of 0.03 R or
lower needed to test these models more definitively.

If the clouds in these two sightlines are in fact ionized by an external
flux of ionizing radiation, then the \ha\ intensity of each H~I cloud is a
measure of the incident ionizing flux.  Specifically, the one-sided
incident flux is $2.2 \times 10^6$ $I_{\mathrm{H}\alpha}$ photons
cm$^{-2}$ s$^{-1}$, for $I_{\mathrm{H}\alpha}$ in Rayleighs and a
temperature of 10$^4$ K (e.g., \citealt{reynolds95, tufte98}).  These
observations therefore imply a variation in the intensity of the ambient
ionizing flux, from less than $1.3 \times 10^5$ photons cm$^{-2}$ s$^{-1}$
at the intermediate velocity (--50~\kms) H~I cloud in the Lockman Window
to about $4.4 \times 10^5$ photons cm$^{-2}$ s$^{-1}$ for the low velocity
clouds.  The \ha\ intensity of the high velocity (--90~\kms) component
toward HD 93521 gives a flux of $5 \times 10^4$ photons cm$^{-2}$
s$^{-1}$; however, because there is no associated H~I detected at this
velocity, the H~II is probably density bounded, and thus the derived value
for the ionizing flux must be considered as a lower limit. If the higher
velocity clouds are farther from the Galactic midplane than the lower
velocity clouds \citep{kuntz96}, these results indicate a decrease in
the ionizing flux with distance above the plane.

\acknowledgements

We thank Mark Quigley for his help with the observations and Trudy
Tilleman for her on-site monitoring of Kitt Peak sky conditions during
these observations.  N. R. H. gratefully acknowledges support through a
Barry M. Goldwater Scholarship from the Excellence in Education
Foundation.  This work, including the operation of WHAM, was funded by
the National Science Foundation through grant AST96-19424.

\clearpage

\clearpage

\begin{deluxetable}{ccccc}
\tablecolumns{5}
\tabletypesize{\scriptsize}
\tablewidth{0pt}
\tablecaption{List of \ha\ Observations\label{table:ha_obs}}
\tablehead{
\colhead{} & \colhead{} & \multicolumn{3}{c}{Total Exposure Time}\\
\colhead{} & \colhead{} & \multicolumn{3}{c}{(s)}\\
\cline{3-5}
\colhead{} & \colhead{} & \colhead{HD 93521} &
\colhead{Lockman Window} &
\colhead{Off A}\\
\colhead{Data Set} & \colhead{Observation Date} & 
\colhead{($\ell = 183\fdg1$, $b = +62\fdg2$)} & 
\colhead{($\ell = 148\fdg5$, $b = +53\fdg0$)} & 
\colhead{($\ell = 163\fdg5$, $b = +53\fdg5$)}}
\startdata
1a & 1997 Nov 5 & 1200 & 1200 & \nodata\\
1b & 1997 Nov 7 & 1200 & 1200 & \nodata\\
1c & 1997 Nov 8 & 600  & 600  & \nodata\\
2a & 1998 Apr 22\tablenotemark{a} & \nodata & 1800 & \nodata\\
2b & 1998 Apr 23\tablenotemark{a} & \nodata & 1800 & \nodata\\
3 & 1999 Jan 18 & 3000 & 6000\tablenotemark{b} & \nodata\\
4a & 1999 Apr 16 & \nodata & 600 & \nodata\\
4b & 1999 Apr 20 & \nodata & 600 & \nodata\\ 
5 & 1999 Dec 12 & \nodata & 3600 & 3600\\
6 & 2000 Feb 2\tablenotemark{a} & \nodata & 3600 & 3600\\
7 & 2000 Feb 4 & 3600 & \nodata & 3600\\
\enddata
\tablenotetext{a}{Spectra from this date have the geocoronal line 
positioned on the blue side of the 200~\kms\ spectral range.}
\tablenotetext{b}{Individual spectra are 1200~s exposures.}
\end{deluxetable}

\clearpage

\begin{deluxetable}{ccccc}
\tablecolumns{5}
\tabletypesize{\scriptsize}
\tablewidth{0pt}
\tablecaption{List of \oiw\ Observations\label{table:oi_obs}}
\tablehead{
\colhead{} & \colhead{} & 
\multicolumn{3}{c}{Total Exposure Time}\\
\colhead{} & \colhead{} & \multicolumn{3}{c}{(s)}\\
\cline{3-5}
\colhead{} & \colhead{} & \colhead{HD 93521} &
\colhead{Lockman Window} &
\colhead{Off A}\\
\colhead{Data Set} & \colhead{Observation Date} & 
\colhead{($\ell = 183\fdg1$, $b = +62\fdg2$)} & 
\colhead{($\ell = 148\fdg5$, $b = +53\fdg0$)} & 
\colhead{($\ell = 163\fdg5$, $b = +53\fdg5$)}
}
\startdata
8a & 1998 Nov 18 & 2400 & 1200 & \nodata\\
8b & 1998 Nov 21 & 2400 & 1200 & \nodata\\
9a & 1999 Jan 18 & 2400  & 2400\tablenotemark{a}  & \nodata\\
9b & 1999 Jan 22 & 2400 & 1200 & \nodata\\
10a & 1999 Apr 16 & 1200 & 600 & \nodata\\
10b & 1999 Apr 20 & 3600 & 1800 & \nodata\\
11 & 2000 Feb 4 & 2400 & \nodata & 2400\\
\enddata
\tablenotetext{a}{Individual spectra are 1200~s exposures.}
\end{deluxetable}

\clearpage

\begin{deluxetable}{cccccc}
\tablecolumns{6}
\tabletypesize{\small}
\tablewidth{0pt}
\tablecaption{Faint Atmospheric Lines near \ha\label{table:atmlines}}
\tablehead{
\colhead{} & \colhead{Wavelength} & \colhead{Width\tablenotemark{b}}
& \colhead{Adopted Intensity} 
& \multicolumn{2}{c}{Measured Intensity\tablenotemark{c}}\\
\cline{5-6}
\colhead{Line\tablenotemark{a}} & \colhead{(\AA)} 
& \colhead{(\kms)} & \colhead{(\% of Line 5 Intensity)} 
& \multicolumn{2}{c}{(R)} 
}
\startdata
1 & 6559.55 & 10 & 54 & 0.070\tablenotemark{e} & \nodata\\ 
2 & 6559.92 & 10 & 24 & 0.052 & \nodata\\ 
3 & 6560.86 & 10 & 20 & 0.041 & \nodata\\ 
4 & 6561.45 & 15 & 33 & 0.057 & 0.027\\ 
5 & 6561.88 & 10 & 100 & 0.13 & 0.11\\ 
6 & 6562.25 & 10 & 30 & 0.033 & \nodata\\ 
7 & 6563.51 & 10 & 30 & 0.033 & 0.021\\ 
7a & 6563.27\tablenotemark{d} & 10 & \nodata & \nodata & 0.038\\ 
8 & 6563.69 & 10 & 65 & 0.088 & 0.045\\ 
9 & 6563.98 & 10 & 39 & 0.048 & 0.058\\ 
10 & 6564.39 & 15 & 160 & \nodata & 0.17\\ 
11 & 6564.90 & 10 & 36 & \nodata & 0.032\\ 
12 & 6565.44 & 10 & 29 & \nodata & 0.027\\ 
\enddata
\tablenotetext{a}{Line numbers correspond to labels in 
Figure~\ref{fig:mlines} and Figure~\ref{fig:elines}.}
\tablenotetext{b}{Fixed widths were adopted.  See discussion in 
\S\ref{sec:atmlines}.}
\tablenotetext{c}{Left column refers to line intensities in 
Figure~\ref{fig:mlines}, and right column refers to line intensities
in Figure~\ref{fig:elines}.}
\tablenotetext{d}{Wavelength was determined from the spectrum in 
Figure~\ref{fig:elines} only.}
\tablenotetext{e}{Intensity in the fit was fixed at 54\% of Line 5
intensity.}
\end{deluxetable}

\clearpage

\begin{deluxetable}{cccc}
\tablecolumns{4}
\tablewidth{0pt}
\tablecaption{\ha\ Results\label{table:ha_results}}
\tablehead{
\colhead{} & \colhead{V$_{\mathrm{LSR}}$} & \colhead{FWHM} & 
\colhead{\emph{I}}\\ 
\colhead{Direction} & \colhead{(\kms)} & \colhead{(\kms)} & \colhead{(R)}}
\startdata
HD 93521 & --10 $\pm$ 4 
& 22 $\pm$ 6 & 0.20 $\pm$ 0.07\\
& --51 $\pm$ 3 & 39 $\pm$ 7 & 0.15 $\pm$ 0.03\\
& --90 $\pm$ 5 & 25\tablenotemark{a} $\pm$ 5 & 0.023 $\pm$ 0.009\\
Lockman Window & --1 $\pm$ 4 
& 30 $\pm$ 5 & 0.20 $\pm$ 0.05\\
& (--54)\tablenotemark{b,c} & (25) & $<$ 0.06\\
Off A & (0) & (25) & $<$ 0.11\\
& (--50) & (25) & $<$ 0.06 \\
\enddata     
\tablenotetext{a}{Fixed at this value during the fitting.}
\tablenotetext{b}{Parentheses denote no \ha\ component detected.}
\tablenotetext{c}{Velocity of the \hi\ emission in the Lockman Window 
spectrum shown in Figure~\ref{fig:main}.}
\end{deluxetable}
     
\clearpage
      
\begin{deluxetable}{ccccccc}
\tablecolumns{7}
\tablewidth{0pt}
\tablecaption{\oiw\ Results\label{table:oi_results}}
\tablehead{
\colhead{} & \colhead{V$_{\mathrm{LSR}}$} & \colhead{FWHM} & 
\colhead{\emph{I}} &
\colhead{\emph{I}(\oi)/\emph{I}(\ha)} &
\multicolumn{2}{c}{\emph{n}(H$^+$)/\emph{n}(H$^{\circ}$)}\\ 
\cline{6-7}
\colhead{Direction} & \colhead{(\kms)} & \colhead{(\kms)} & \colhead{(R)} &
\colhead{(energy units)} & \colhead{6000 K} & \colhead{10,000 K}}
\startdata
HD 93521 & (--51)\tablenotemark{a,b} & (34)\tablenotemark{c} & 
$<$0.060 & $<$0.43 &
$>$0.13 & $>$1.4\\ 
& (--51) & (34) & 0.026\tablenotemark{d} & 0.18 & 0.32 & 3.3\\
\enddata     
\tablenotetext{a}{Parentheses denote no component detected.}
\tablenotetext{b}{Velocity of the \ha\ emission component.}
\tablenotetext{c}{Calculated based on the width of the \ha\ component
and a gas temperature of 8000 K.}
\tablenotetext{d}{Intensity of the gaussian fit in Figure~\ref{fig:oi}.}
\end{deluxetable}

\end{document}